\documentclass{aa}

\usepackage{graphicx}
\usepackage{txfonts}
\usepackage{natbib}     
\usepackage{xcolor} 
\usepackage{threeparttable} 
\usepackage[colorlinks=true,citecolor=blue, linkcolor=blue]{hyperref} 

\bibpunct{(}{)}{;}{a}{}{,} 

\begin{document}

   \title{Detection of the hydrogen Balmer lines in the ultra-hot Jupiter WASP-33b}

   \author{F. Yan\inst{1}
   		\and
   		A. Wyttenbach\inst{2}
   		\and
   		N. Casasayas-Barris\inst{3,4}
   		\and
   		A. Reiners\inst{1}
          \and
          E. Pall\'e\inst{3,4}
          \and
          Th. Henning\inst{5}
          \and
          P.~Molli\`ere\inst{5}
          \and
          S.~Czesla\inst{6}
          \and
          L.~Nortmann\inst{1}
          \and
          K. Molaverdikhani\inst{7,5}
          \and
          G.~Chen\inst{8}
          \and
          I.~A.~G.~Snellen\inst{9}
          \and
           M.~Zechmeister\inst{1}
          \and
          C.~Huang\inst{10}
   		\and  
   		I.~Ribas\inst{11,12},  A.~Quirrenbach\inst{7}, J.~A.~Caballero\inst{13}, P.~J.~Amado\inst{14}
   		\and
                D.~Cont\inst{1}, S.~Khalafinejad\inst{7}, J.~Khaimova\inst{1}, M.~L\'opez-Puertas\inst{14}, 
		D. Montes\inst{15}, E.~Nagel\inst{16}, M.~Oshagh\inst{3,4}, S.~Pedraz\inst{17}
		\and  
		M.~Stangret\inst{3,4}
             }
                
\institute{Institut f\"ur Astrophysik, Georg-August-Universit\"at, Friedrich-Hund-Platz 1, 37077 G\"ottingen, Germany\\
	\email{fei.yan@uni-goettingen.de}
\and Universit\'e Grenoble Alpes, CNRS, IPAG, 38000 Grenoble, France
                \and
	Instituto de Astrof{\'i}sica de Canarias (IAC), Calle V{\'i}a Lactea s/n, 38200 La Laguna, Tenerife, Spain
\and
Departamento de Astrof{\'i}sica, Universidad de La Laguna, 38026  La Laguna, Tenerife, Spain
\and
Max-Planck-Institut f{\"u}r Astronomie, K{\"o}nigstuhl 17, 69117 Heidelberg, Germany
\and
Hamburger Sternwarte, Universit{\"a}t Hamburg, Gojenbergsweg 112, 21029 Hamburg, Germany
\and
Landessternwarte, Zentrum f\"ur Astronomie der Universit\"at Heidelberg, K\"onigstuhl 12, 69117 Heidelberg, Germany
\and
Key Laboratory of Planetary Sciences, Purple Mountain Observatory, Chinese Academy of Sciences, Nanjing 210023, China
\and
	Leiden Observatory, Leiden University, Postbus 9513, 2300 RA, Leiden, The Netherlands
\and	
	Lunar and Planetary Laboratory, 1629 E. University Blvd., P.O. Box 210092, Tucson, AZ 85721-0092, United States of America
\and
Institut de Ci\`encies de l'Espai (CSIC-IEEC), Campus UAB, c/ de Can Magrans s/n, 08193 Bellaterra, Barcelona, Spain
\and
Institut d'Estudis Espacials de Catalunya (IEEC), 08034 Barcelona, Spain
\and
Centro de Astrobiolog{\'i}a (CSIC-INTA), ESAC, Camino bajo del castillo s/n, 28692 Villanueva de la Ca{\~n}ada, Madrid, Spain
\and
Instituto de Astrof{\'i}sica de Andaluc{\'i}a (IAA-CSIC), Glorieta de la Astronom{\'i}a s/n, 18008 Granada, Spain
\and
Departamento de F\'{i}sica de la Tierra y Astrof\'{i}sica 
and IPARCOS-UCM (Instituto de F\'{i}sica de Part\'{i}culas y del Cosmos de la UCM), 
Facultad de Ciencias F\'{i}sicas, Universidad Complutense de Madrid, E-28040, Madrid, Spain
\and
Th{\"u}ringer Landessternwarte Tautenburg, Sternwarte 5, 07778 Tautenburg, Germany
\and
Centro Astron{\'o}nomico Hispano-Alem{\'a}n (CSIC-MPG), Observatorio Astron{\'o}nomico de Calar Alto, Sierra de los Filabres, E-04550 G{\'e}rgal, Almer\'ia, Spain
\\      }
        \date{Received August 31, 2020; accepted November 15, 2020}


  \abstract
{ Ultra-hot Jupiters (UHJs) are highly irradiated giant exoplanets with extremely high day-side temperatures, which lead to thermal dissociation of most of the molecular species.
It is expected that the neutral hydrogen atom is one of the main species in the upper atmospheres of ultra-hot Jupiters. Neutral hydrogen has been detected in several UHJs by observing its Balmer line absorption.
Here, we report four transit observations of the ultra-hot Jupiter WASP-33b, performed with the CARMENES and HARPS-North spectrographs, and the detection of the H$\alpha$, H$\beta$, and H$\gamma$ lines in the planetary transmission spectrum.
The combined H$\alpha$ transmission spectrum of the four transits has an absorption depth of 0.99$\pm$0.05\,\%, which corresponds to an effective radius of 1.31$\pm$0.01 $R_\mathrm{p}$. The strong H$\alpha$ absorption indicates that the line probes the high-altitude thermosphere.
We further fitted the three Balmer lines using the \texttt{PAWN} model, assuming that the atmosphere is hydrodynamic and in LTE. We retrieved a thermosphere temperature $12200^{+1300}_{-1000}$\,K and a mass-loss rate ${\rm \dot{M}}=10^{11.8^{+0.6}_{-0.5}}$ g\,s$^{-1}$. The retrieved large mass-loss rate is compatible with the ``Balmer-driven'' atmospheric escape scenario, in which the stellar Balmer continua radiation in the near-ultraviolet is substantially absorbed by the excited hydrogen atoms in the planetary thermosphere.
 }

   \maketitle

%

\section{Introduction}
Ultra-hot Jupiters (UHJs) are the hottest giant exoplanets and they are extensively irradiated by their host stars. These planets are ideal laboratories to study the chemistry and physics of planetary atmospheres under extreme conditions.
Theoretical modelling of UHJ atmospheres \citep[e.g.,][]{Lothringer2018,Parmentier2018,Kitzmann2018,Helling2019} suggests that the day-sides as well as the terminators of UHJs are extremely hot and probably dominated by atoms and ions instead of molecules due to thermal dissociation and ionisation. 

The thermal emission spectra of several UHJs, including HAT-P-7b \citep{Mansfield2018}, WASP-12b \citep{Stevenson2014}, WASP-18b \citep{Arcangeli2018}, and WASP-103b \citep{Kreidberg2018}, have been observed with the \textit{Hubble Space Telescope} (\textit{HST}). These thermal spectra exhibit a lack of $\mathrm{H_2O}$ features, which is probably due to thermal dissociation \citep{Parmentier2018}.
On the other hand, emission spectroscopy with high-resolution spectrographs has revealed the existence of neutral Fe in KELT-9 \citep{Pino2020}, WASP-189b \citep{Yan2020}, and WASP-33b \citep{Nugroho2020W33}.
In addition to the thermal emission spectra, phase curve observations have been performed for several UHJs, including WASP-33b \citep{Zhang2018, Essen2020}, WASP-121b \citep{Daylan2019, Bourrier2020-TESS}, and KELT-9b \citep{Wong2020,Mansfield2020}. These observations suggest that these UHJs have relatively low day-night temperature contrasts and relatively high heat transport efficiencies. 
The increased heat transport efficiency in UHJs could be explained by a new physical mechanism -- thermal dissociation and recombination of $\mathrm{H_2}$ \citep{Bell2018, Komacek2018}.

Transmission spectroscopy has also been widely used in probing the atmospheres of UHJs, and various atomic and ionic species have been detected. In the atmosphere of KELT-9b -- the hottest exoplanet discovered so far, the hydrogen Balmer lines as well as multiple metal lines (including \ion{Fe}{i}, \ion{Fe}{ii}, \ion{Ti}{ii}, \ion{Mg}{i}, and \ion{Ca}{ii}) have been detected \citep{Yan2018, Hoeijmakers2018, Cauley2019, Hoeijmakers2019, Yan2019,Turner2020}. 
Various metals as well as the Balmer lines have been detected in KELT-20b/MASCARA-2b \citep{Casasayas-Barris2018,Casasayas-Barris2019, Stangret2020, Nugroho2020, Hoeijmakers2020}. 
\ion{Ca}{ii} is detected in WASP-33b \citep{Yan2019}.
The Balmer lines and metals including \ion{Na}{i}, \ion{Mg}{ii}, \ion{Fe}{i}, \ion{Fe}{ii}, \ion{Cr}{i}, and \ion{V}{i} have been detected in WASP-121b \citep{Sing2019, Bourrier2020, Gibson2020,Cabot2020,Ben-Yami2020}. 
$\mathrm{H\alpha}$ and \ion{Mg}{ii} have been discovered in WASP-12b \citep{Fossati2010, Jensen2018}.
Neutral Fe has also been detected at the terminator of WASP-76b \citep{Ehrenreich2020}.

Planets experiencing strong stellar irradiation are thought to undergo hydrodynamic atmospheric escape \citep[e.g.,][and references therein]{Owen2019}. The hydrodynamic escape in hydrogen-dominated atmospheres is normally driven by the absorption of stellar extreme-ultraviolet (EUV) flux \citep{Yelle2004,Tian2005,Salz2016}.
However, \cite{Fossati2018} found that heating due to atomic absorption of the stellar UV and optical flux drives the atmospheric escape of UHJs orbiting early-type stars. \cite{Garcia-Munoz2019} further proposed that the absorption of the hydrogen Balmer line series can enhance and even drive the atmospheric escape of UHJs orbiting hot stars.

Observations of atmospheric escape have been performed with the hydrogen $\mathrm{Ly{\alpha}}$ line \citep{Vidal-Madjar2003, Etangs2012, Ehrenreich2015} as well as metal lines in the ultraviolet \citep{Fossati2010, Sing2019, Cubillos2020}, using the STIS spectrograph on \textit{HST}.
The helium 10833 $\mathrm{\AA}$ line has recently been used in probing escaping atmosphere of planets orbiting active stars \citep[e.g.,][]{Spake2018, Nortmann2018, Allart2018, Salz2018, Lampon2020, Palle2020}.

The hydrogen Balmer lines can also be used to probe high-altitude atmospheres and study atmospheric escape. For example, \cite{Yan2018} estimated the Jeans escape rate of KELT-9b with the $\mathrm{H{\alpha}}$ absorption line. Recently, \citet{Wyttenbach2020} modelled the Balmer lines with a hydrodynamic model and retrieved the mass-loss rate of KELT-9b. 
The Balmer lines have been detected in four UHJs so far: KELT-9b \citep{Yan2018,Cauley2019,Turner2020}, KELT-20b \citep{Casasayas-Barris2018, Casasayas-Barris2019}, WASP-12b \citep{Jensen2018}, and WASP-121b \citep{Cabot2020}. 
Besides, the $\mathrm{H{\alpha}}$ line has been detected in two non-UHJ planets -- HD 189733b \citep{Jensen2012, Barnes2016, Cauley2016, Cauley2017, Cauley2017-HD189} and WASP-52b \citep{Chen2020}. The $\mathrm{H{\alpha}}$ absorption in these two planets probably originates from the excitation of hydrogen atoms due to stellar $\mathrm{Ly{\alpha}}$ line and Lyman continuum irradiation \citep{Christie2013,Huang2017}.

Here, we present the discovery of the Balmer line absorption during the transit of WASP-33b -- a UHJ (equilibrium temperature $\sim$ 2710 K) orbiting an A5 star \citep{Cameron2010}. Several species have previously been detected in its planetary atmosphere, including TiO \citep{Haynes2015,Nugroho2017}, \ion{Ca}{ii} \citep{Yan2019}, \ion{Fe}{i} \citep{Nugroho2020W33}, and evidence of AlO \citep{Essen2019} and FeH \citep{Kesseli2020}.
The paper is organised as follows. We describe the transit observations and data analysis in Sect. 2. The observational results are presented in Sect. 3. In Sect. 4, we present the hydrodynamic model of the Balmer lines and discuss the atmospheric escape of WASP-33b. The conclusions are summarised in Sect. 5.

%
\section{Data and analysis}

\subsection{Observations}
We observed four transits of \object{WASP-33b} with two spectrographs. The observation logs are summarised in Table \ref{obs_log}.

Two transits were observed with the CARMENES \citep{Quirrenbach2018} installed at the 3.5 m telescope of the Calar Alto Observatory on 5 January 2017 and 16 January 2017. The visual channel of the CARMENES spectrograph has a resolution of \textit{R} $\sim$ 94\,600 and a wavelength coverage of 520--960\,nm.
The first night was photometric (i.e., ideal weather condition during the observation) and the second night was partially cloudy.

Another two transits were observed with the HARPS-North (HARPS-N) spectrograph mounted on the Telescopio Nazionale Galileo telescope on 17 October 2018 and 8 November 2018. The instrument has a resolution of \textit{R} $\sim$ 115\,000 and a wavelength coverage of 383--690\,nm. We used the order-merged one-dimensional spectra from the HARPS-N pipeline (Data Reduction Software). The spectra have an over-sampled wavelength step of 0.01 $\mathrm{\AA}$. We re-binned the spectrum every 3 wavelength points by averaging so that each wavelength point corresponds to 0.03 $\mathrm{\AA}$, which is similar to the CARMENES pixel size at the $\mathrm{H\alpha}$ line centre (0.030 $\mathrm{\AA}$).
Both nights were photometric. However, the spectral flux from the first night observation had a large drop when the telescope was pointing close to the zenith. Such a phenomenon also occurred during transit observations in \cite{Casasayas-Barris2019}, which was probably caused by a problem with the atmospheric dispersion corrector (ADC).

\begin{figure}
   \centering
   \includegraphics[width=0.5\textwidth]{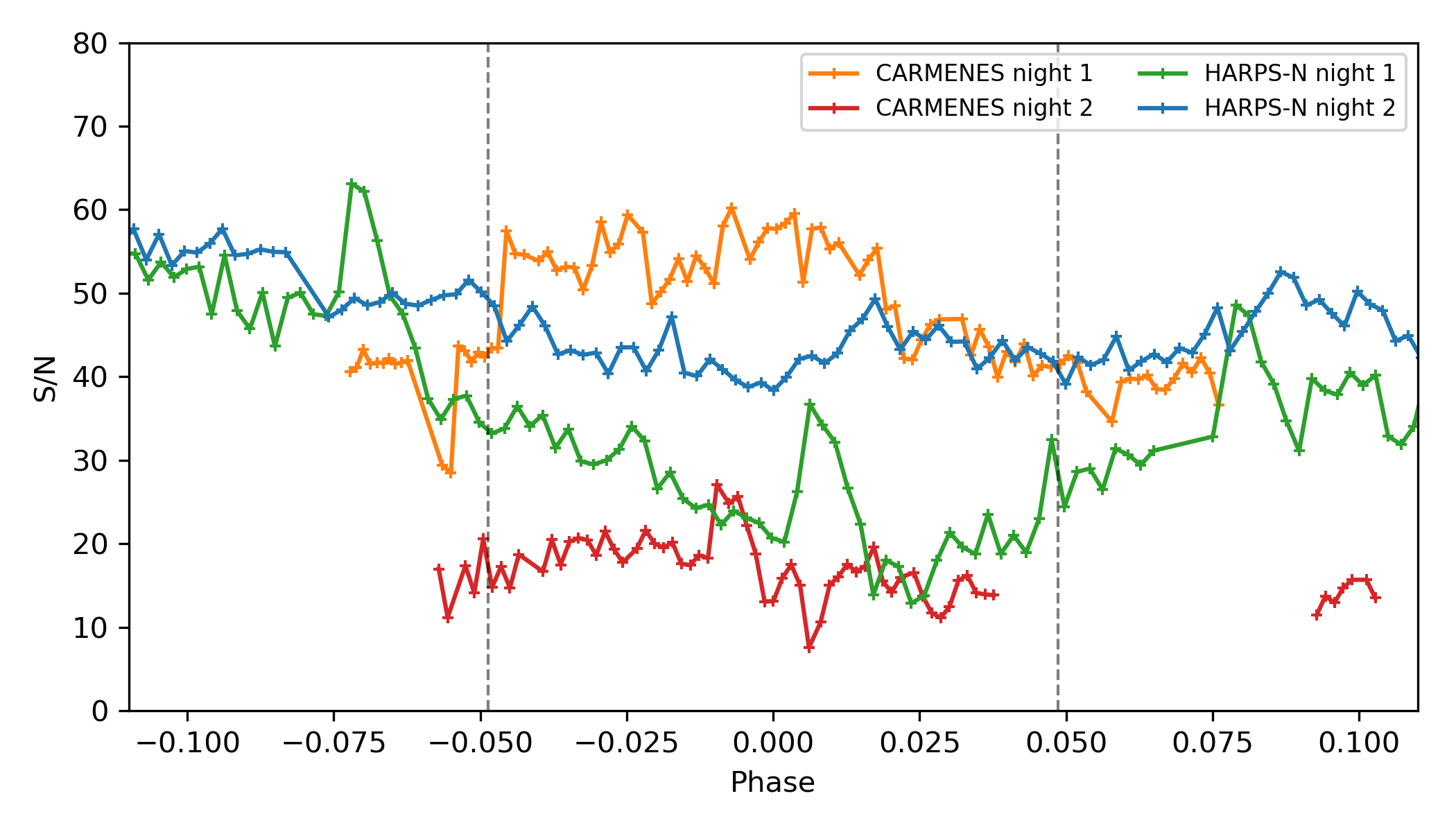}
      \caption{Signal-to-noise ratio per wavelength point ($\sim 0.03 \mathrm{\AA}$) at the $\mathrm{H\alpha}$ line centre. The dashed lines indicate the beginning and end of transit.
      }
         \label{SNR}
   \end{figure}

The signal-to-noise ratio (S/N) per wavelength point ($\sim$ 0.03 $\mathrm{\AA}$) at the $\mathrm{H\alpha}$ line centre is plotted in Fig.\ref{SNR}. 
Among the four transits, night-1 from CARMENES and night-2 from HARPS-N observations have much higher S/N and, therefore, were used in \cite{Yan2019} for the detection of ionised calcium. In this work, we use and combine all four transits.

%
\begin{table*}
\caption{Observation logs.}             
\label{obs_log}      
\centering                          
\begin{threeparttable}
	\begin{tabular}{l c c c c c c c}        
	\hline\hline \noalign{\smallskip}                
 	Instrument &  & 	Date & Observing Time (UT) & Airmass change & Exposure time [s] & $N_\mathrm{spectra}$ \\     
	\hline \noalign{\smallskip}                      
 CARMENES	& Night-1 & 2017-01-05	 & 19:28--23:49 & 1.00--1.54 & 120 \tablefootmark{a} & 93 \\ 
CARMENES & Night-2 & 2017-01-16	 & 19:25--00:07 & 1.01--2.03 & 120 & 66 \\ 
	\hline \noalign{\smallskip}
HARPS-N & Night-1 & 2018-10-17	 & 21:39--05:46 & 1.64--1.01--1.54 & 200 & 124 \\          		
HARPS-N & Night-2 & 2018-11-08	 & 19:59--05:01 & 1.74--1.01--1.87 & 200 & 141 \\          		
\hline \noalign{\smallskip}                                  
	\end{tabular}
	\tablefoot{
\tablefoottext{a}{The first 19 spectra had exposure time below 120 s. }
}
\end{threeparttable}      
\end{table*}

~\\
\subsection{Obtaining the transmission spectral matrix}
We investigated the $\mathrm{H\alpha}$ line (6562.79$\,\mathrm{\AA}$) using both the CARMENES and HARPS-N observations and the $\mathrm{H\beta}$ (4861.35$\,\mathrm{\AA}$) and $\mathrm{H\gamma}$ (4340.47$\,\mathrm{\AA}$) lines from the HARPS-N observations. The data reduction method is similar to the method in \cite{Yan2018}.

The spectra were first normalised and shifted into the Earth's rest frame. We then removed the telluric absorption lines using a theoretical transmission spectral model of the Earth's atmosphere \citep{Yan2015b}.
The spectra were subsequently aligned into the stellar rest frame by correcting the barycentric radial velocity and the stellar systemic velocity \citep[-3.0$\,\mathrm{km\,s^{-1}}$, ][]{Nugroho2017}. We obtained an out-of-transit master spectrum by adding up all the out-of-transit spectra with the squared S/N as weight. We then divided each spectrum by the master spectrum in order to remove the stellar lines.
The residual spectrum was subsequently filtered with a Gaussian high-pass filter ($\sigma \sim$ 300 points) to remove large scale features on the continuum spectrum, which may be attributed to the stability of the HARPS-N ADC, stellar pulsation, or the imperfect normalisation of the blaze variation.

We combined the two CARMENES observations as well as the two HARPS-N observations by binning the spectra with an orbital phase step of 0.005. The binning was performed by averaging the spectra within each phase bin with the squared S/N as weight.

\begin{figure*}
   \centering
   \includegraphics[width=0.90\textwidth]{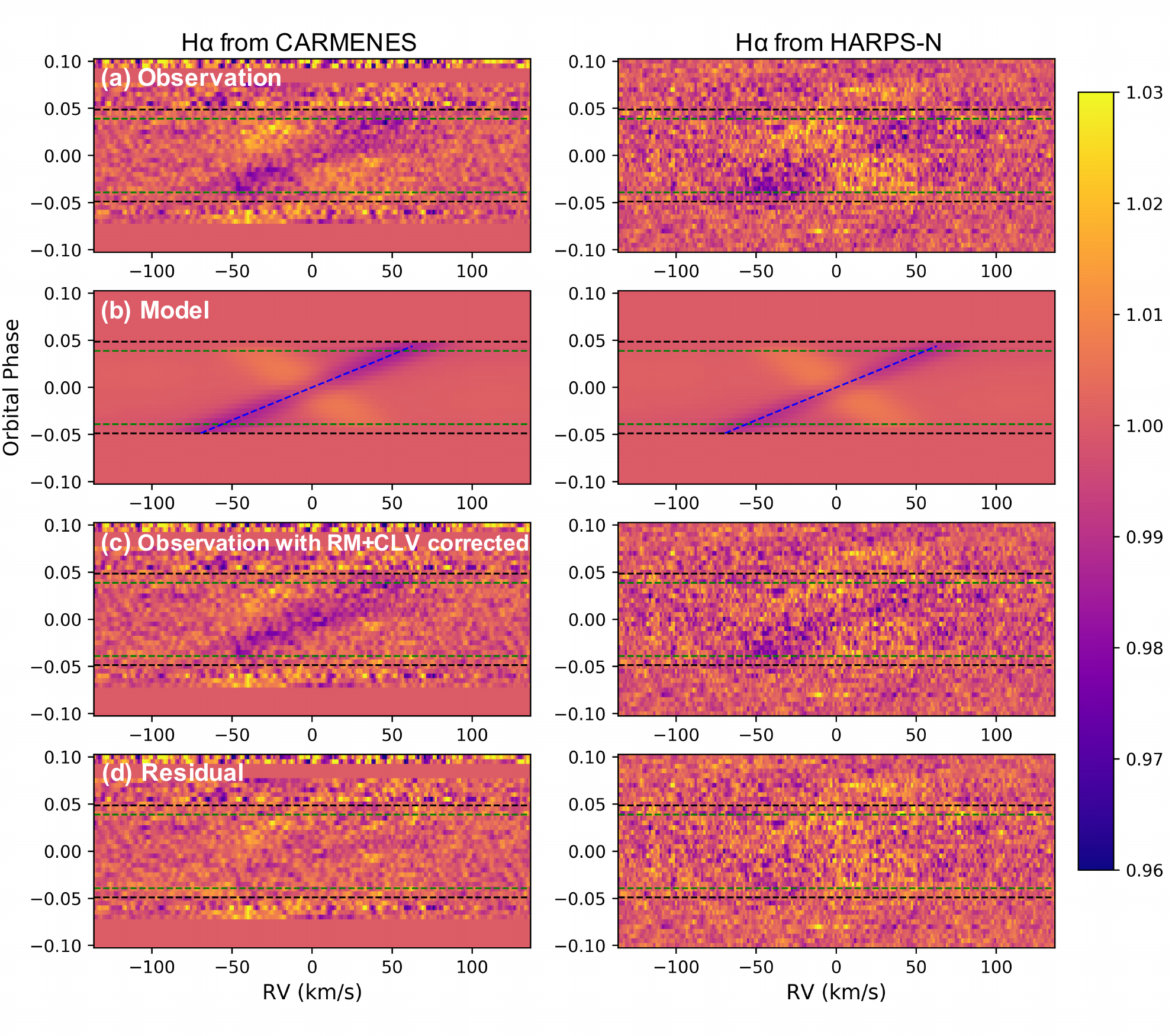}
      \caption{Transmission spectral matrices for the $\mathrm{H\alpha}$ line from the CARMENES observations (left) and the HARPS-N observations (right). The color bar indicates the value of relative flux.
\textit{(a)} The observed transmission spectra. The x axis is wavelength expressed in RV relatively to the $\mathrm{H\alpha}$ line centre (6562.79 $\mathrm{\AA}$) in the stellar rest frame. The horizontal dashed lines indicate the four contacts of transit.
\textit{(b)} The best-fit model from the MCMC analysis. The model includes the $\mathrm{H\alpha}$ transmission spectrum and the stellar line profile change (i.e. the CLV and RM effects). 
The blue dashed line indicates the RV of the planetary orbital motion plus a constant shift ($V_\mathrm{centre}$). Although the models extend into the ingress and egress regions on the matrices, the fit was only performed on the fully in-transit data.
\textit{(c)} The observed transmission spectra with the RM and CLV effects corrected.
\textit{(d)} The residual between the observation and the model.}
         \label{Ha-CAR+HAR-map}
   \end{figure*}

\begin{figure*}
   \centering
   \includegraphics[width=0.90\textwidth]{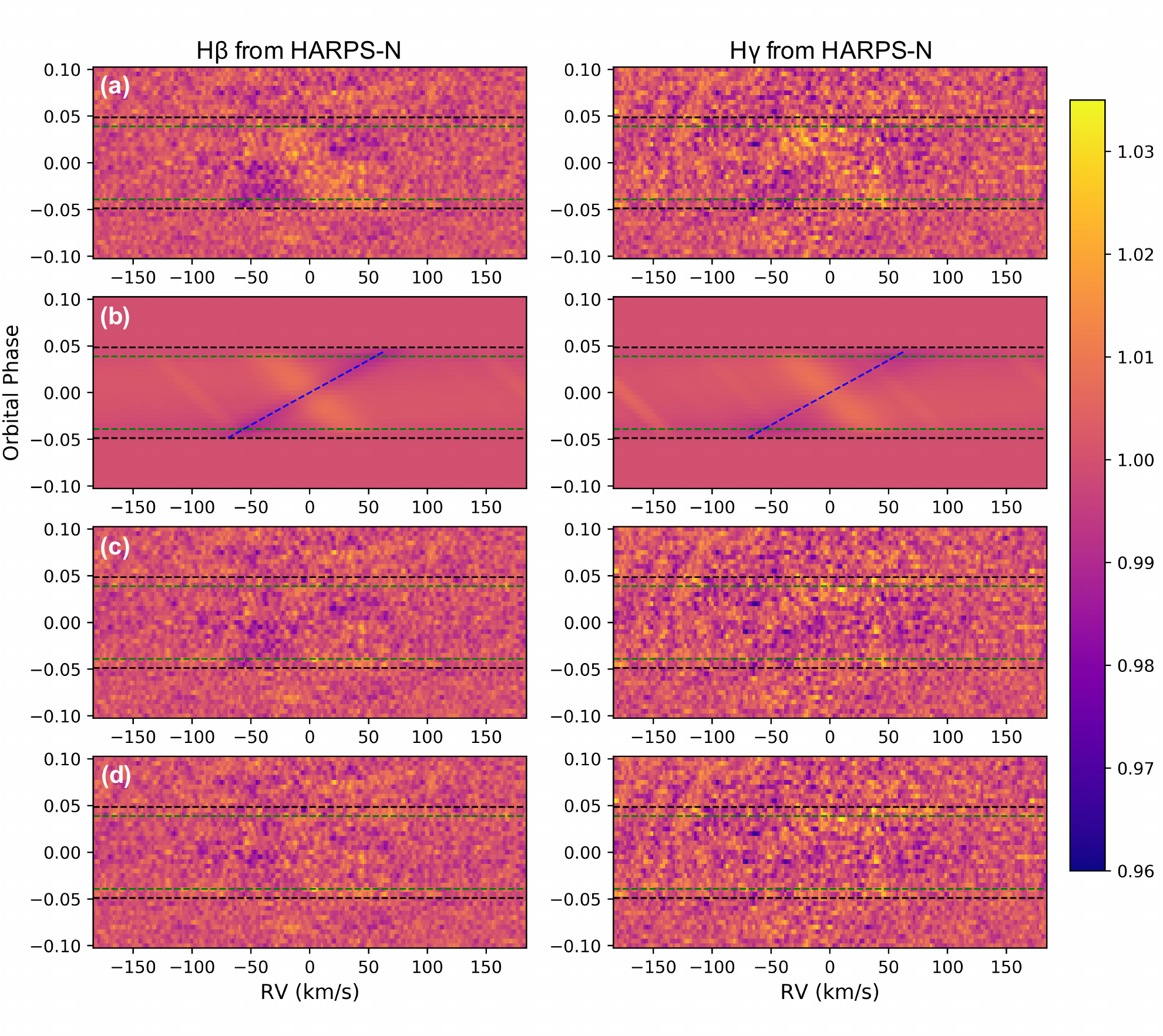}
      \caption{Same as Fig.~\ref{Ha-CAR+HAR-map} but for the $\mathrm{H\beta}$ line (left) and the $\mathrm{H\gamma}$ line (right) from the HARPS-N observations.
      }
         \label{Hb+Hc}
   \end{figure*}

By applying these procedures, we obtained a transmission spectral matrix for each of the $\mathrm{H\alpha}$, $\mathrm{H\beta}$, and $\mathrm{H\gamma}$ lines from the HARPS-N observations and an $\mathrm{H\alpha}$ spectral matrix from the CARMENES observations (upper panels in Figs.~\ref{Ha-CAR+HAR-map} and \ref{Hb+Hc}).

%

%

\subsection{Model of stellar RM and CLV effects}
The stellar line profile varies during the transit due to the Rossiter-McLaughlin (RM) effect \citep{Queloz2000} and the centre-to-limb variation (CLV) effect \citep{Yan2015a, Czesla2015, Yan2017}. We modelled the RM and CLV effects simultaneously following the method described in \cite{Yan2018}. We used the same stellar and planetary parameters as in \cite{Yan2019}.

The planetary orbit of WASP-33b undergoes a nodal precession \citep{Johnson2015, Iorio2016, Watanabe2020}. We adopted the orbital change rates from \cite{Johnson2015} and calculated the expected orbital inclination ($i$) and spin-orbit angle ($\lambda$) at the dates of our observations. Because the observation dates are very close for the two CARMENES transits as well as for the two HARPS-N transits, the changes of the orbital parameters between them are negligible. Therefore, we set $i = 89.50$ deg and $\lambda = -114.05$ deg for the combined CARMENES transmission matrix; $i = 90.14$ deg and $\lambda = -114.93$ deg for the combined HARPS-N transmission matrix.

\subsection{Fitting the observed spectral matrix}
We fitted the observed transmission spectral matrix with a model consisting of two components: the planetary absorption and the stellar line profile change.
We assumed that the planetary absorption has a Gaussian profile described by the full width at half maximum (FWHM), the absorption depth ($h$), and the radial velocity (RV) shift of the observed line centre compared to the theoretical value ($V_\mathrm{centre}$). 
The semi-amplitude of the planetary orbital motion ($K_\mathrm{p}$) is fixed to the expected $K_\mathrm{p}$ value ($231\pm3$  $\mathrm{km\,s^{-1}}$), which is calculated with the planetary orbital parameters.
The stellar line profile change caused by the RM and CLV effects is fixed to the results as calculated in Sect. 2.3.

We sampled from the posterior probability distribution using the Markov Chain Monte Carlo (MCMC) simulations with the \texttt{emcee} tool \citep{Mackey2013}. 
We only used the fully in-transit data (i.e. excluding the ingress and egress phases).

\section{Results and discussion}

\subsection{$\mathrm{H\alpha}$ transmission spectrum}
\begin{figure}
   \centering
   \includegraphics[width=0.5\textwidth]{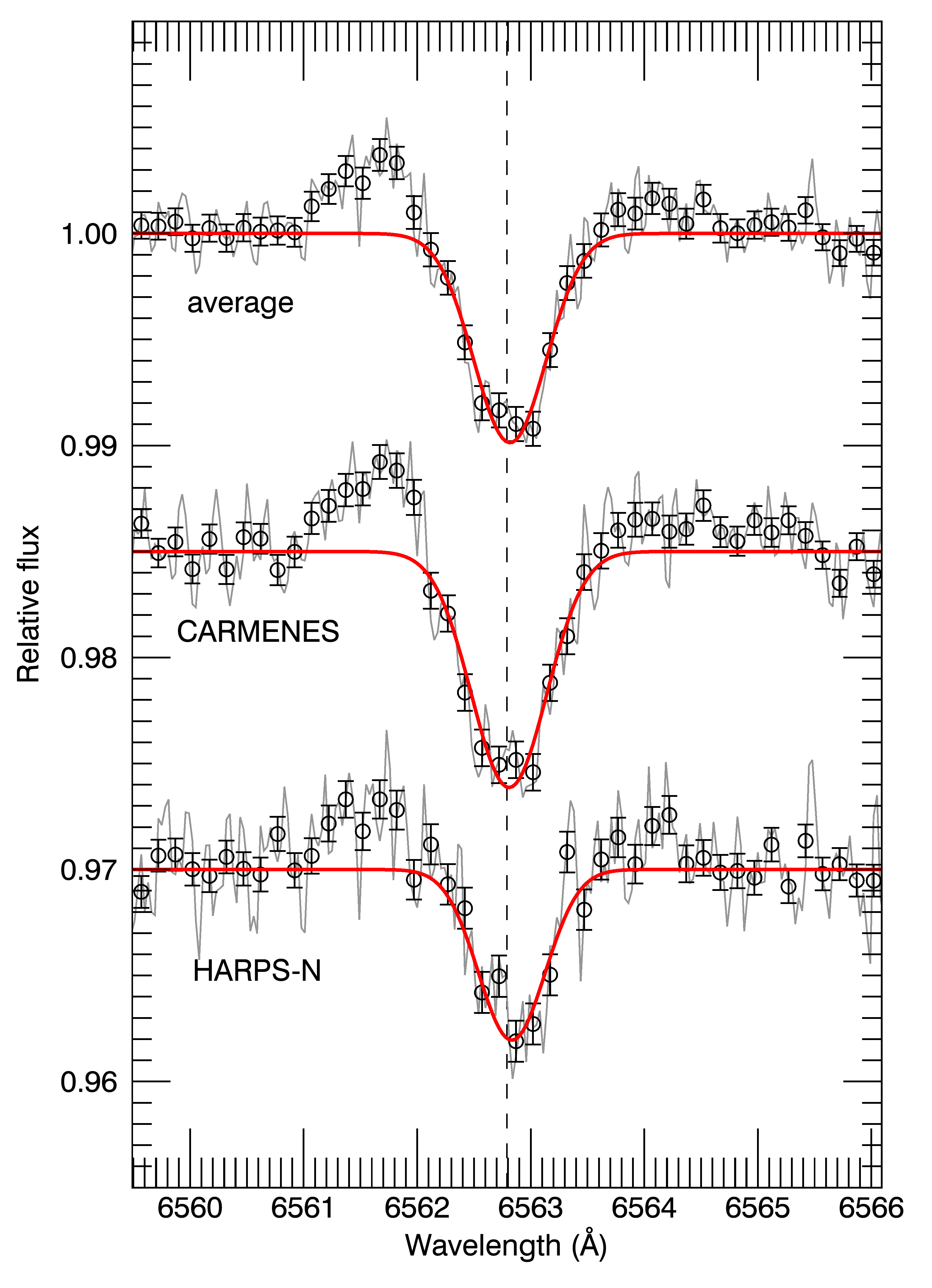}
      \caption{Transmission spectra of the $\mathrm{H\alpha}$ line. The black circles are spectra binned every five points ($\sim$ 0.15 $\AA$) and the grey lines are the original spectra (i.e., $\sim$ 0.03 $\AA$ per point). The red lines are the best-fit Gaussian functions. The vertical dashed line indicates the rest wavelength line centre. The CLV and RM effects are corrected.
      An offset of the y-axis is applied to the spectra for clarity.
      }
         \label{Spec-Ha}
   \end{figure}

The transmission spectral matrices are shown in Fig.~\ref{Ha-CAR+HAR-map}a. The $\mathrm{H\alpha}$ absorption is clearly detected in both CARMENES and HARPS-N data. The best-fit models of the planetary absorption feature as well as the stellar CLV and RM effects are shown in Fig.~\ref{Ha-CAR+HAR-map}b, and the best-fit parameters are summarised in Table \ref{Tab-fit-reuslt-tran}.
In order to obtain the one-dimensional transmission spectra, we firstly corrected the CLV and RM effects. Then, the residual spectra were shifted into the planetary rest frame. We subsequently averaged all the fully in-transit spectra to derive the final one-dimensional transmission spectra, which are presented in Fig.~\ref{Spec-Ha}. The $\mathrm{H\alpha}$ transmission spectrum of each night is shown in Fig.~\ref{App-Ha-individual}.

The obtained FWHM is 31.6$_{-3.6}^{+4.1}$ $\mathrm{km\,s^{-1}}$ for HARPS-N and 35.6$_{-2.0}^{+2.2}$ $\mathrm{km\,s^{-1}}$ for CARMENES. These values are smaller than those of KELT-9b \citep{Yan2018, Cauley2019, Turner2020} while slightly higher than those of KELT-20b \citep{Casasayas-Barris2018, Casasayas-Barris2019}. The large FWHM indicates that the $\mathrm{H\alpha}$ absorption is optically thick \citep{Huang2017}.

The measured RV shift of the line centre is  2.0$\pm$1.9 $\mathrm{km\,s^{-1}}$ for HARPS-N and 0.8$\pm$1.1 $\mathrm{km\,s^{-1}}$ for CARMENES. The RV shift has been used to measure high-altitude winds at the planetary terminator \citep{Snellen2010, Wyttenbach2015,Louden2015,Brogi2016}.
Nevertheless, the measured $V_\mathrm{centre}$ is relative to the stellar systemic RV and there is a large uncertainty of the stellar RV of WASP-33. This is because precisely measuring the absolute RV of fast-rotating A-type stars is intrinsically challenging. For example, the reported systemic RVs of WASP-33 deviate by several $\mathrm{km\,s^{-1}}$ \citep{Cameron2010, Lehmann2015, Nugroho2017, Cauley2020-W33}. Therefore, we conclude that we do not detect any significant winds at the terminator of WASP-33b considering the uncertainties in the measured $V_\mathrm{centre}$ values and the stellar systemic RV.

\begin{figure}
   \centering
   \includegraphics[width=0.5\textwidth, height=0.3\textwidth]{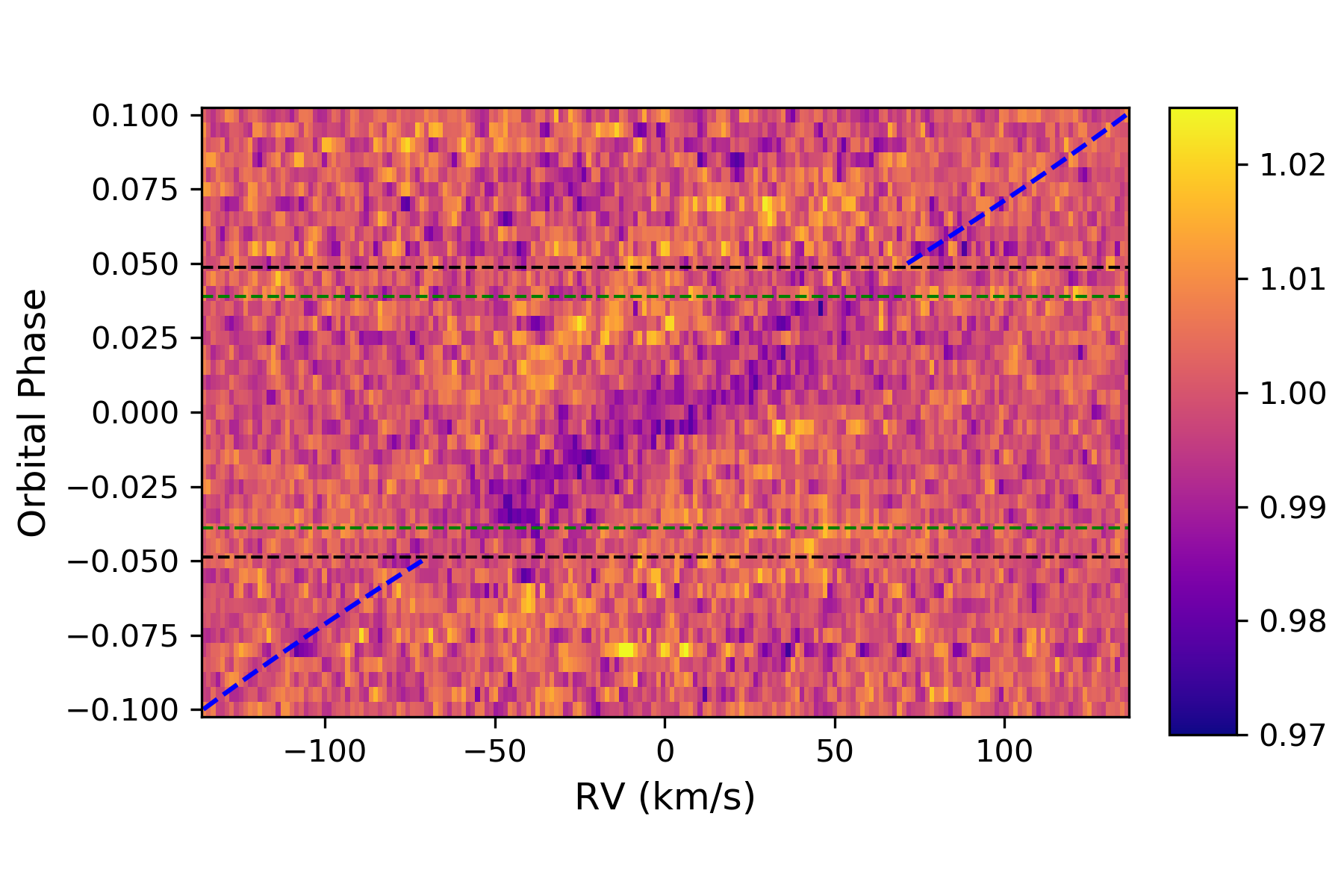}
      \caption{Combined transmission spectral matrix of the CARMENES and HARPS-N results for the $\mathrm{H\alpha}$ line. The stellar line profile change due to the CLV and RM effects has been removed before averaging. The horizontal dashed lines indicate the four contacts of transit and the diagonal dashed lines denote the planetary orbital RV.
      }
         \label{Ha-combine}
   \end{figure}

\begin{figure}
   \centering
   \includegraphics[width=0.49\textwidth]{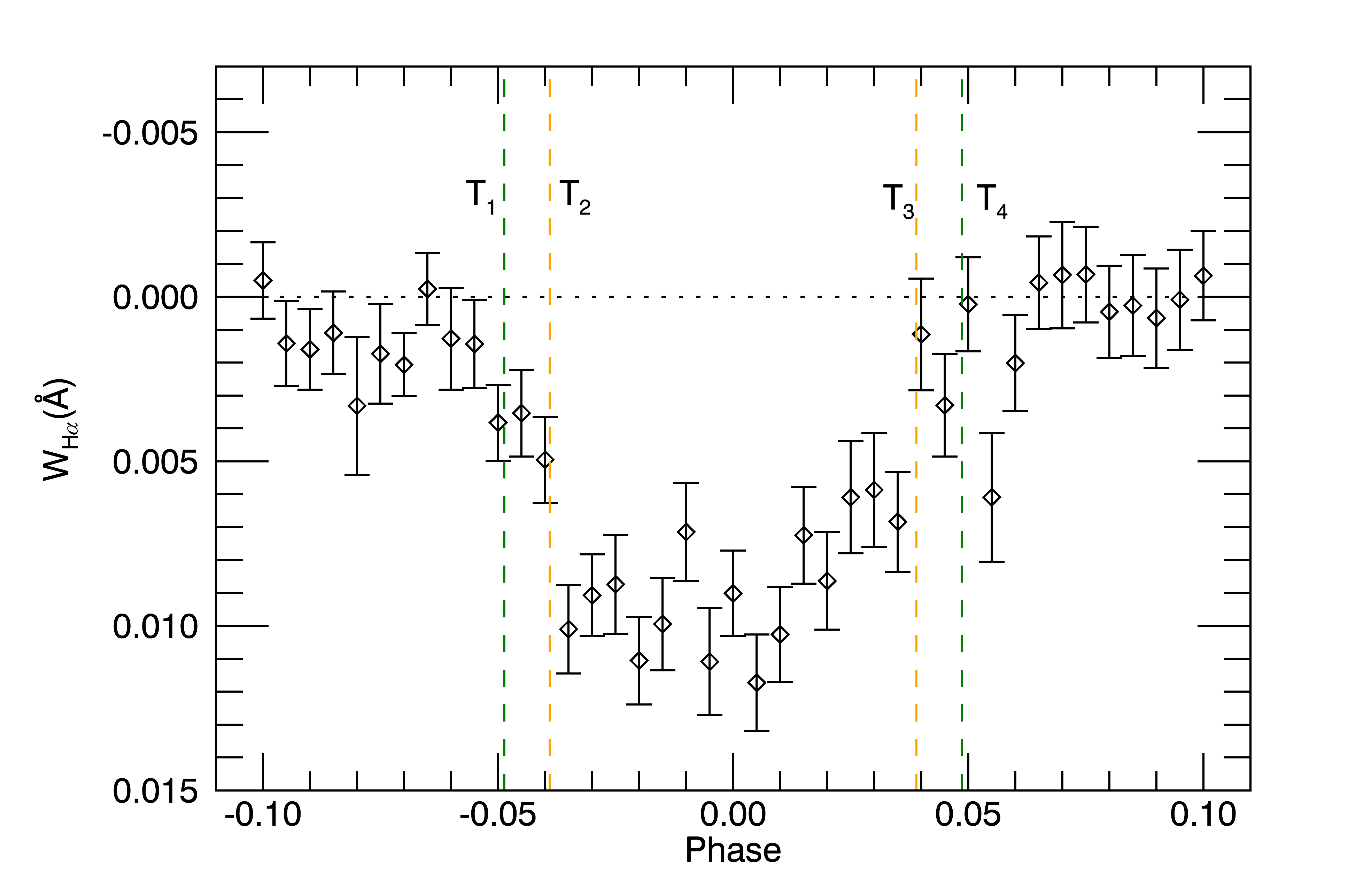}
      \caption{Time series of the $\mathrm{H\alpha}$ equivalent width. The values are measured on the combined transmission spectral matrix in Fig.~\ref{Ha-combine}. The vertical dashed lines indicate the first ($\mathrm{T_1}$), second ($\mathrm{T_2}$), third ($\mathrm{T_3}$), and fourth ($\mathrm{T_4}$) contacts of the transit. The horizontal line denotes $W_\mathrm{H\alpha}$ = 0.
      }
         \label{LC-Ha}
   \end{figure}

We further combined the CARMENES and HARPS-N transmission spectral matrices using the binning method as described in Section 2.2. The stellar CLV and RM effects were already corrected before the averaging. The combined matrix is presented in Fig.~\ref{Ha-combine} and the best-fit parameters are listed in Table \ref{Tab-fit-reuslt-tran}.
We calculated the equivalent width of the absorption line ($W_\mathrm{H\alpha}$) using the same method as in \cite{Yan2018}, except that the integration range was set as $\pm 35\,\mathrm{km\,s^{-1}}$ to match the observed FWHM. Fig.~\ref{LC-Ha} shows the time series of $W_\mathrm{H\alpha}$. There is no obvious pre- or after-transit absorption, although the absorption depth is slightly stronger during the first-half transit.

In general, the fitted parameters between CARMENES and HARPS-N are consistent. However, the CARMENES $\mathrm{H\alpha}$ absorption is somewhat stronger than the HARPS-N absorption. Such a slight difference between the two instruments is also observed for the $\mathrm{H\alpha}$ line in KELT-20b \citep{Casasayas-Barris2019} and KELT-9b \citep{Yan2018,Wyttenbach2020,Turner2020}. 
Although the difference could be from random variations, there may also be systematic residuals, which could be due to instrumental effects (e.g., non-linearity, the stability of the HARPS-N ADC) or data reduction procedures (e.g., imperfect normalisation, removal of stellar and telluric lines). 

For the case of WASP-33b, the slight difference between the CARMENES and HARPS-N results could be caused by the stellar pulsations, which can affect the stellar $\mathrm{H\alpha}$ line profile.
The host star is a known $\mathrm{\delta}$ Scuti star with pronounced pulsations \citep{Cameron2010, Essen2014, Kovacs2013}.
For the CARMENES spectrum in Fig.~\ref{Spec-Ha}, there is a bump feature on the left of the planetary absorption line. A weaker bump feature is also present in the HARPS-N spectrum.
Such a bump feature is also observed in the transmission spectrum of the \ion{Ca}{ii} infrared triplet lines \citep{Yan2019}, which are obtained using the same transit data as in this work.
On the combined transmission spectral matrix in Fig.~\ref{Ha-combine}, there are bright stripes on the left and right sides of the planetary absorption signal and these stripes extend beyond the transit. These stripes are probably the stellar pulsation signatures, which generate the bump features as observed on the transmission spectra in Fig.~\ref{Spec-Ha}.
For individual CARMENES and HARPS-N observation, the position and strength of the pulsation features were not the same during the transits. Therefore, the stellar pulsation could introduce a difference between the CARMENES and HARPS-N results.
In a preprint posted while the present paper was under review, \cite{Cauley2020-W33} reported the detection of the Balmer lines in WASP-33b using the PEPSI spectrograph mounted on the Large Binocular Telescope. 
Their obtained absorption line strength is quantitatively different but generally consistent with our CARMENES and HARPS-N results, considering possible effects of the stellar pulsation.

\begin{figure}
   \centering
   \includegraphics[width=0.49\textwidth]{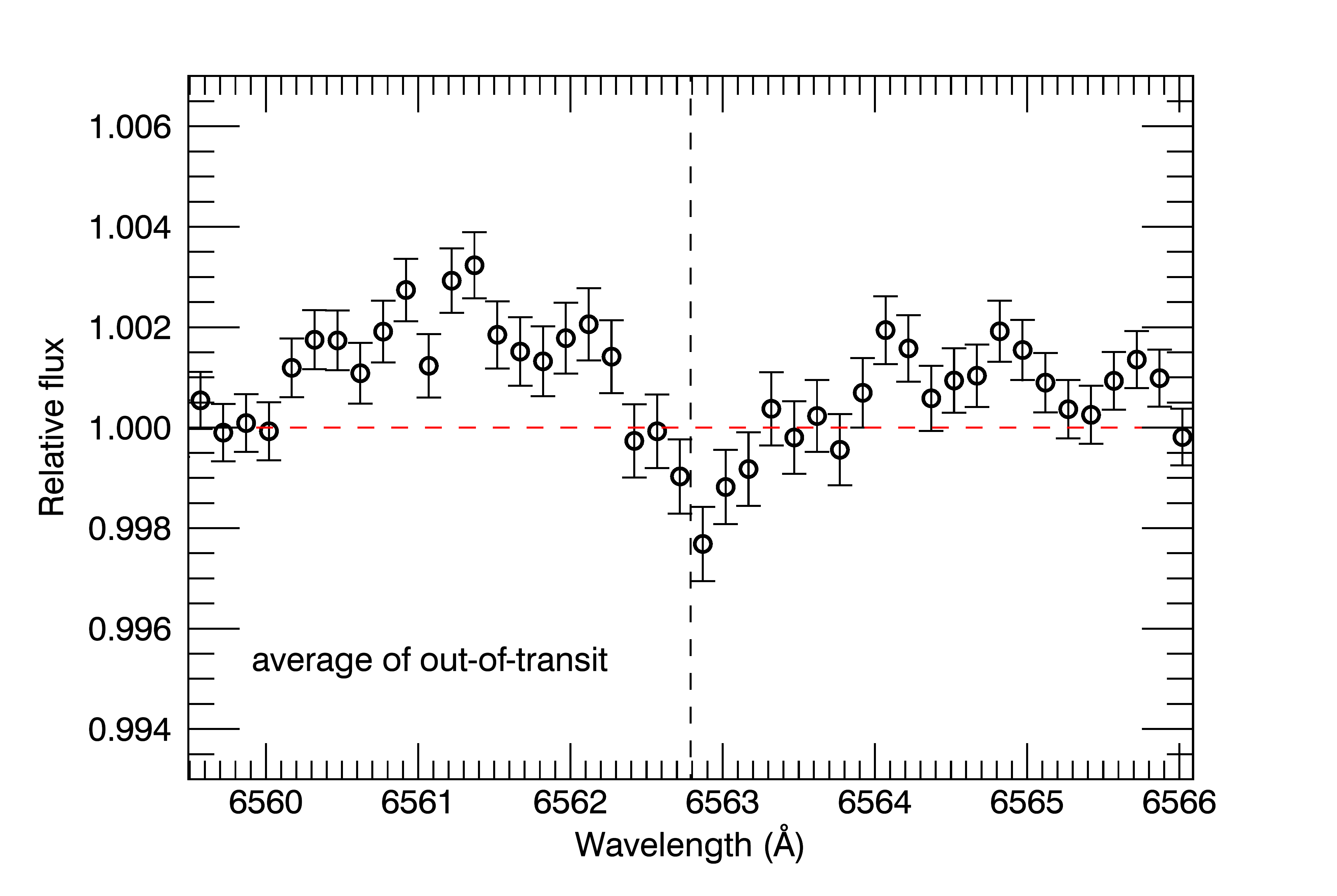}
      \caption{Average out-of-transit spectrum of the $\mathrm{H\alpha}$ transmission spectral matrix in Fig.\ref{Ha-combine}. The vertical dashed line denotes the line centre. These spectral features likely originate from the stellar pulsation.
      }
         \label{spec-OOT}
   \end{figure}

Although the detection of the $\mathrm{H\alpha}$ line is unambiguous, its strength is potentially affected by the stellar pulsation. To correct the effect of the pulsation, a detailed analysis of the variation in the stellar Balmer lines is required. Such an analysis requires data taken with high S/N and is beyond the scope of this paper. However, considering that the pulsating periods are not synchronous with the planetary orbital motion \citep{Essen2014}, the pulsating contribution to the transmission spectrum should be statistically reduced when combining the four transit spectra together. To evaluate the effect of pulsation, we combined the out-of-transit spectra on the spectral matrix in Fig.\ref{Ha-combine} (i.e., phases --0.10 to --0.05 and  +0.05 to +0.10) and assumed the in-transit orbital velocity to repeat during out-of-transit. The obtained out-of-transit spectrum (Fig.\ref{spec-OOT}) shows ripple-like features with semi-amplitude of $\sim$0.2\%, which are most likely the results of the stellar pulsation. The effect of the pulsation to the $\mathrm{H\alpha}$ transmission spectrum should be at a similar order of these ripple features.

We note that \cite{Valyavin2018} analysed the transit light curve of WASP-33b observed with the $\mathrm{H\alpha}$ filter. These latter authors found that the $\mathrm{H\alpha}$ transit depth is significantly deeper than the transit depths measured in broad bands. 
Since we detect the strong $\mathrm{H\alpha}$ absorption line with high-resolution spectroscopy, we confirm that the photometric result of \cite{Valyavin2018} is evidence of the $\mathrm{H\alpha}$ absorption in the planetary atmosphere.

%

%

%

\subsection{$\mathrm{H\beta}$ and $\mathrm{H\gamma}$ transmission spectrum}
\begin{figure}
   \centering
   \includegraphics[width=0.5\textwidth]{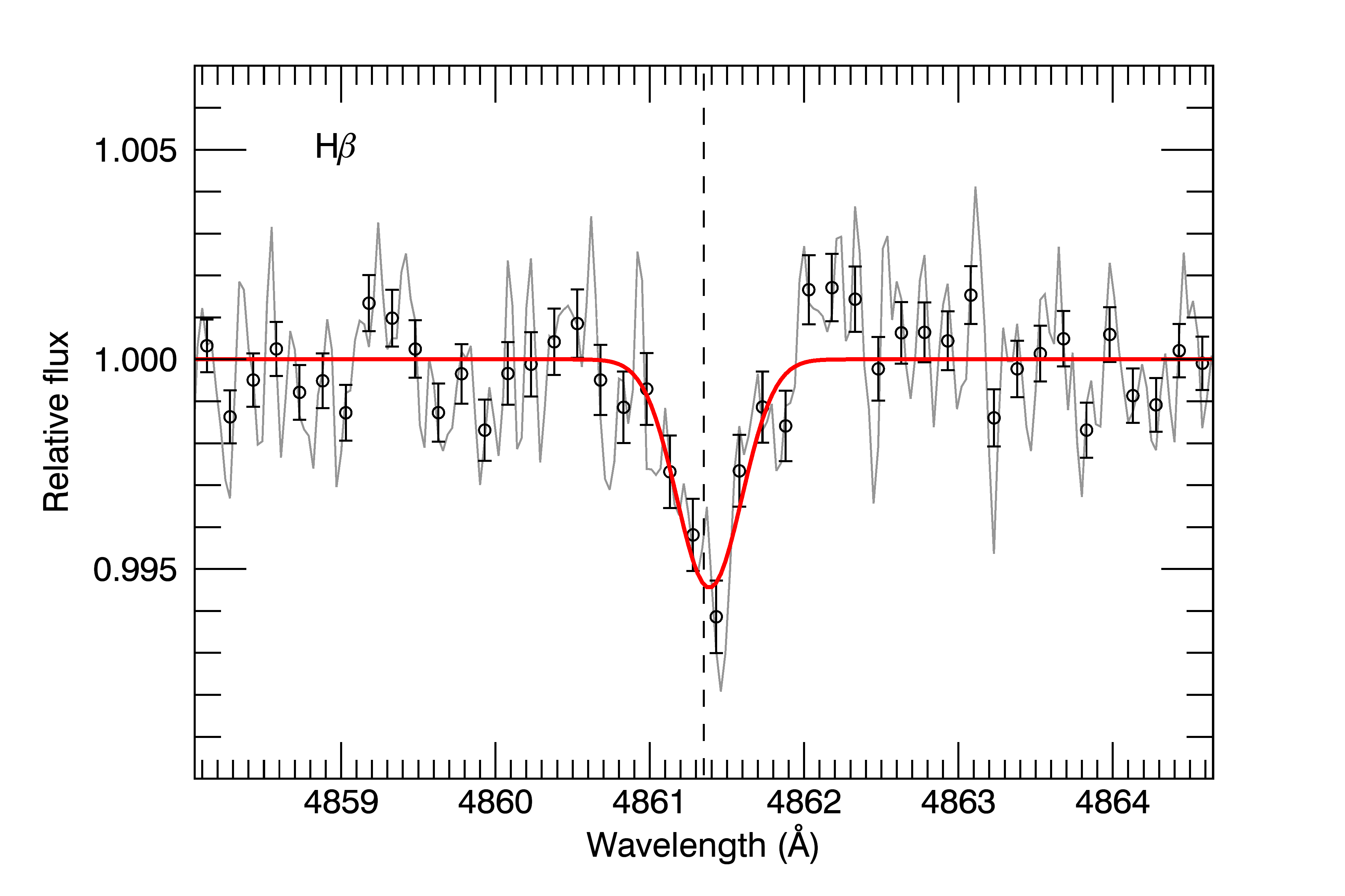}
   \includegraphics[width=0.5\textwidth]{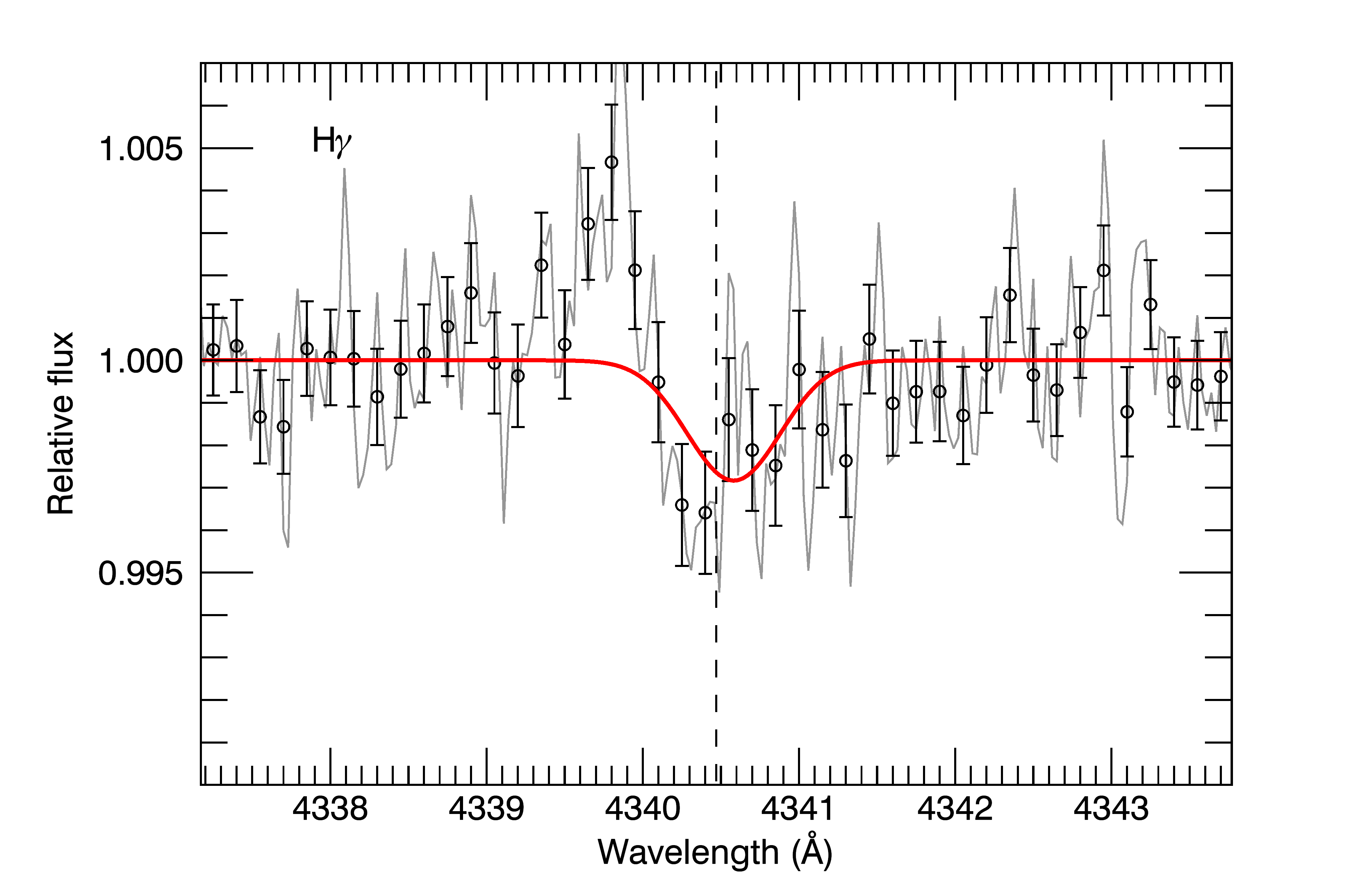}
      \caption{Transmission spectra of the $\mathrm{H\beta}$ line (upper panel) and the $\mathrm{H\gamma}$ line (lower panel). 
      }
         \label{Spec-Hb+Hc}
   \end{figure}

The $\mathrm{H\beta}$ and $\mathrm{H\gamma}$ lines are only covered by the HARPS-N spectrograph. The absorption signals are relatively weak compared to the $\mathrm{H\alpha}$ line. The best-fit parameters are shown in Table \ref{Tab-fit-reuslt-tran} with the final transmission spectra in Fig.~\ref{Spec-Hb+Hc}. The detection of $\mathrm{H\beta}$ is clear while the $\mathrm{H\gamma}$ signal is less prominent. 
The line depth of the $\mathrm{H\beta}$ absorption is smaller than that of the $\mathrm{H\alpha}$ line, but their FWHM values are relatively similar to each other. This is also the case for the $\mathrm{H\alpha}$ and $\mathrm{H\beta}$ lines in KELT-9b \citep{Cauley2019,Wyttenbach2020} and KELT-20b \citep{Casasayas-Barris2019}.

%

%
\begin{table*}
\caption{Fit results of the transmission spectral matrices.}             
\label{Tab-fit-reuslt-tran}     
\centering   
\begin{tabular}{l l c c c c c}        
\hline\hline\noalign{\smallskip}                 
    ~ & ~ & $V_\mathrm{centre}$ [$\mathrm{km\,s^{-1}}$] & FWHM [$\mathrm{km\,s^{-1}}$]   & Line depth [\%]  & $R_\mathrm{eff}$ [$R_\mathrm{p}$] & Detection significance\\ 	
	\hline \noalign{\smallskip}                      
	~ & CARMENES &  0.8$\pm$1.1 & 35.6$_{-2.0}^{+2.2}$ & 1.11$\pm$0.07 & 1.34$\pm$0.02 & 16 $\mathrm{\sigma}$\\ 
	$\mathrm{H\alpha}$ & HARPS-N &  2.0$\pm$1.9 & 31.6$_{-3.6}^{+4.1}$ &  0.81$\pm$0.09 & 1.26$\pm$0.03 & 9 $\mathrm{\sigma}$\\
	~ & combination  & 1.2$\pm$0.9 & 34.3$\pm$1.6 & 0.99$\pm$0.05 & 1.31$\pm$0.01 & 20 $\mathrm{\sigma}$\\
	\hline  \noalign{\smallskip}                             
	$\mathrm{H\beta}$ & HARPS-N &  2.2$\pm$1.7 & 30.6$_{-4.3}^{+4.9}$ & 0.54$\pm$0.07 & 1.18$\pm$0.02 & 8 $\mathrm{\sigma}$\\
	\hline  \noalign{\smallskip}                             
	$\mathrm{H\gamma}$ & HARPS-N & 7$\pm$15 & 49$_{-26}^{+25}$ &  0.28$_{-0.15}^{+0.09}$ & 1.10$_{-0.05}^{+0.03}$ & 2.3 $\mathrm{\sigma}$\\
\hline                                         
\end{tabular}    
\end{table*}

\subsection{Model of the Balmer lines}

\subsubsection{Estimation of the atmospheric conditions}
In order to obtain a rough estimate of the possible atmospheric conditions, and before modelling the lines, one can use the \citet{Lecavelier2008} formula to estimate the atmospheric scale-height in the region probed by the Balmer lines. Indeed, the altitude of absorption $z$ is proportional to the hydrogen Balmer line oscillator strengths $\ln(gf)$: $\Delta z = H\Delta \ln(gf)$, where $H=k_BT/\mu g$ is the pressure scale-height. The $\ln(gf)$ values are 1.635, -0.046, and -1.029 for the H$\alpha$, H$\beta$, and H$\gamma$ lines, respectively.
Taking into account all our measurements from CARMENES and HARPS-N, we computed  $H=9\,200\pm1\,200$~km. Considering the decrease of the gravity $g$ with altitude, we estimated that $T/\mu\simeq15\,100\pm2\,100$~[K/u] (at $z\sim1.2$\,${\rm R_P}$). As it is likely that molecular hydrogen is dissociated under these conditions, we can further assume that $\mu$ is between 0.66 and 1.26 (the atmosphere is dominated by a mixture of ionised and neutral hydrogen and helium). Hence, we estimated the upper atmosphere temperature to be between 8\,600 and 21\,700 K. The lower end of the $T$ range should be preferred since when the temperature increases, the amount of ionised hydrogen increases, making $\mu$ decreases as well.

\subsubsection{Model set-up}
To interpret the observed Balmer lines in WASP-33b, we employed the \texttt{PAWN} model (PArker Winds and Saha-BoltzmanN atmospheric model) developed by \citet{Wyttenbach2020}. This tool is a 1-D model of an exoplanet upper atmosphere linked to an MCMC retrieval algorithm. Its purpose is to retrieve parameters of the thermosphere regions (e.g., the temperature, mass-loss rate) from high-resolution transmission spectra. Key features of the \texttt{PAWN} model are summarised here.
First, we can choose the atmospheric structure to be hydrostatic (barometric law) or hydrodynamic (Parker wind transonic solution), with the base density or the mass-loss rate being a free parameter, respectively. The atmospheric profile is assumed to be isothermal in both cases, with the temperature being an additional free parameter. We also assume the atmosphere to be in chemical equilibrium, with Solar abundances. We use a chemical grid calculated with the equilibrium chemistry code presented in \citet{Molliere2017}, from which we interpolate the volume mixing ratios and other useful quantities according to the atmospheric structure.
As we detected Balmer lines, we focus on the neutral hydrogen. In local thermodynamic equilibrium (LTE), the number densities of the different electronic states follow the Boltzmann distribution. 
Then, the opacities have a Voigt profile and follow the prescriptions of \cite{Kurucz1979,Kurucz1992,Sharp2007}. Finally, the transmission spectrum is computed following \citet{Molliere2019}. The line profiles are broadened taking into account the planetary rotation (tidally locked solid body rotation perpendicular to the orbital plane). The model is also convolved, binned and normalised in order to be comparable to the data.

On top of the atmospheric model parameters presented above, each line centre is a free parameter. For other planetary parameters (e.g., mass and radius), we used the same values as presented in Table 2 of \cite{Yan2019}. For every MCMC chain, we used 10 walkers for each parameter during 2500 steps, with a burn-in size of 500 steps. For each parameter, we used a uniform or log-uniform prior. For the mass-loss rate we put a lower boundary for the prior at $\log_{10}({\rm \dot{M}}$ [g\,s$^{-1}$]) = 9, as it is expected that WASP-33b is undergoing strong atmospheric escape \citep{Fossati2018}. We tried to fit hydrostatic and hydrodynamic structures to see if one structure would be preferred. The Bayesian information criterion (BIC) allows us to compare the results of different models, and to choose the best-fitting model.

\subsubsection{Model results}
We performed MCMC chains on each individual Balmer absorption line from HARPS-N and CARMENES. We also fitted the three Balmer lines simultaneously, using the combined HARPS-N and CARMENES result. It is important to note that since the depth of the H$\alpha$ line is not the same for the two instruments, we would expect some differences in the retrieved parameters.

\begin{figure*}
\centering
\includegraphics[width=0.97\textwidth]{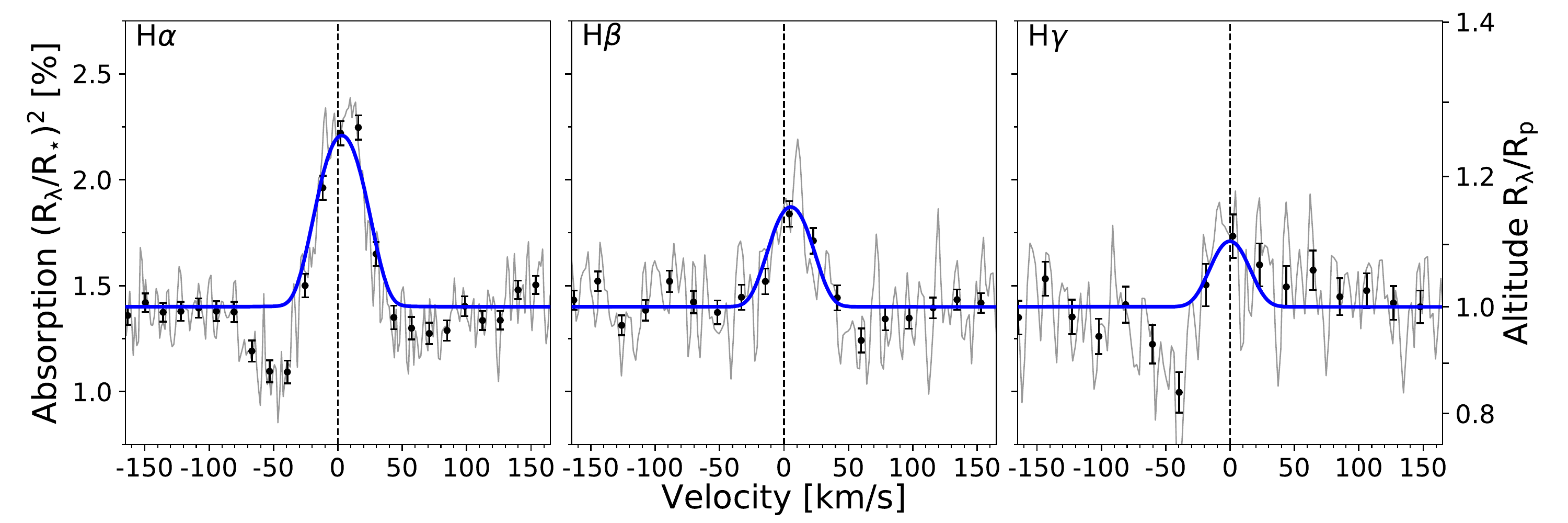}
  \caption{Best-fit \texttt{PAWN} models (blue lines) and the observed transmission spectra (grey lines and black points) in the planetary rest frame. The \texttt{PAWN} models are for the case of a hydrodynamically expanding atmosphere in LTE.
  }
\label{MCMC-PW-LTE-TS}
\end{figure*}

\begin{figure}
\centering
\includegraphics[width=0.5\textwidth]{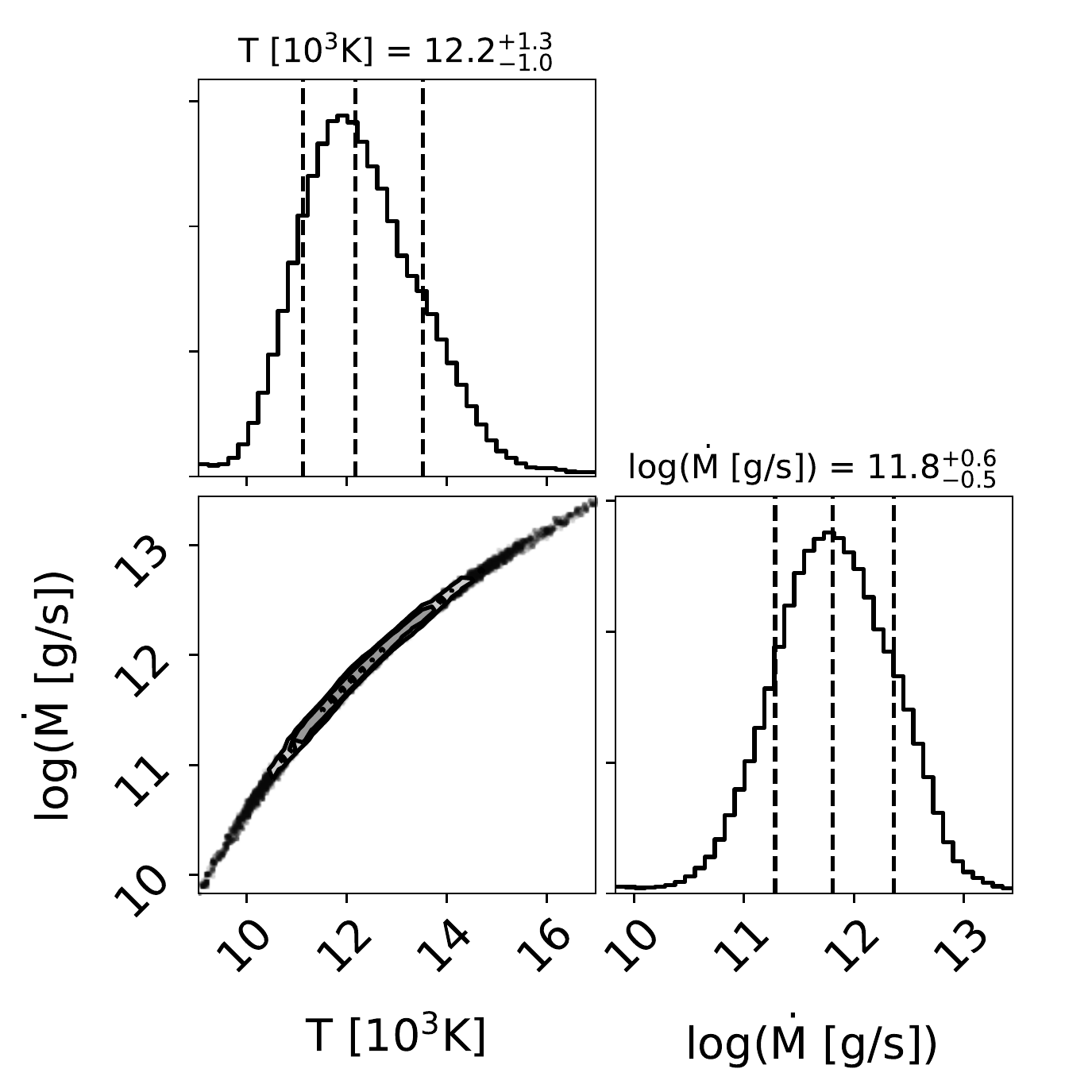}
  \caption{Correlation diagram of the MCMC posterior distributions in the case of a hydrodynamic atmosphere in LTE. The result is for the combined fit of all Balmer lines from HARPS-N and CARMENES. The retrieved parameters are the thermospheric temperature ($T=12\,200^{+1300}_{-1000}$ K) and the atmospheric mass-loss rate (${\rm \dot{M}}=10^{11.8^{+0.6}_{-0.5}}$ g\,s$^{-1}$).
  }
     \label{MCMC-PW-LTE}
\end{figure}

The results from the MCMC model fitting are summarised in Table~\ref{Tab-PAWN-MCMC} for the case of a hydrodynamic atmosphere in LTE. For each detection, the retrieved parameters are compatible. The combined fit (all Balmer lines from HARPS-N and CARMENES) points toward a thermospheric temperature of $T=12\,200^{+1300}_{-1000}$ K and a mass-loss rate of ${\rm \dot{M}}=10^{11.8^{+0.6}_{-0.5}}$ g\,s$^{-1}$. The best-fit spectra and the correlation diagram of the combined fit are presented in Fig.~\ref{MCMC-PW-LTE-TS} and Fig.~\ref{MCMC-PW-LTE}, respectively. Before interpreting this result, we mention here that for each scenario (line or instrument), the absorption line was fitted equally well by a hydrostatic structure ($\Delta$BIC<1). This is because, for a hot Jupiter, it is often possible to find a very similar atmospheric structure for both cases, especially when the temperature is high \citep{Wyttenbach2020}. Nevertheless, an evaporating scenario could be preferred for WASP-33b as suggested by forward modelling \citep{Fossati2018}. This latter study predicted a mass-loss rate of about $10^{11}$ g\,s$^{-1}$ for WASP-33b, which is well in line with our retrieved mass-loss rate.

\begin{table}
\caption{MCMC results of the \texttt{PAWN} modeling for an atmosphere in hydrodynamic expansion and in LTE.}             
\centering   
\begin{tabular}{l l c c}        
\hline\hline \noalign{\smallskip}                
    ~ & ~ & $T$ [$10^3$K] & $\log_{10}({\rm \dot{M}}$ [g\,s$^{-1}$])\\ 	
	\hline \noalign{\smallskip}                      
	$\mathrm{H\alpha}$ & CARMENES &  14.6$_{-2.1}^{+2.4}$ & 12.8$_{-0.8}^{+0.6}$\\ 
	\hline \noalign{\smallskip}                      
	$\mathrm{H\alpha}$ & HARPS-N & 12.6$_{-2.6}^{+4.0}$ & 11.8$_{-1.4}^{+1.3}$\\
	$\mathrm{H\beta}$ & HARPS-N & 12.8$_{-3.3}^{+3.8}$ & 12.1$_{-2.0}^{+1.3}$\\
	$\mathrm{H\gamma}$ & HARPS-N & 12.7$_{-3.2}^{+4.7}$ & 12.0$_{-1.9}^{+1.6}$\\
\hline \noalign{\smallskip}                                  
	All & Combination & 12.2$_{-1.0}^{+1.3}$ & 11.8$_{-0.5}^{+0.6}$\\
	\hline \noalign{\smallskip}                      
\label{Tab-PAWN-MCMC}
\end{tabular}    
\end{table}

Our retrieved mass-loss rate of ${\rm \dot{M}}=10^{11.8^{+0.6}_{-0.5}}$ g\,s$^{-1}$ is close to the maximum energy-limited mass-loss rate \citep{Fossati2018}. This could suggest that the heating efficiency is high (on the order of 10-100\,\%), meaning that most of the irradiation energy goes into expansion ($PdV$ work) and escape. However, according to \citet{Salz2016}, when a hot Jupiter has a relatively high gravitational potential (such as WASP-33b), the heating efficiency should decrease by several orders of magnitude. This hints that the energy-limited computation of \citet{Fossati2018} may not be complete and that some energy sources are not taken into consideration. Indeed, \citet{Garcia-Munoz2019} suggested that hot Jupiters orbiting early type stars could undergo a ``Balmer-driven'' evaporation. This mechanism has been proposed for the ultra-hot Jupiter KELT-9b and is supported by the observations of the Balmer series in its thermosphere \citep{Yan2018,Wyttenbach2020}. This ``Balmer-driven'' mechanism takes place when a sufficient quantity of excited hydrogen is present in the thermosphere and the planet is irradiated with intense stellar near-ultraviolet radiation. In that case, the energy absorbed in the thermosphere from the stellar Balmer irradiation exceeds the one absorbed from the stellar high-energy EUV irradiation. Thus, the thermosphere undergoes a stronger heating and expansion, leading to a higher mass-loss rate, even if the heating efficiency stays moderate. The measured mass-loss rate for WASP-33b is ${\rm \dot{M}}=10^{11.8^{+0.6}_{-0.5}}$ g\,s$^{-1}$, while that of KELT-9b is ${\rm \dot{M}}=10^{12.8\pm0.3}$ g\,s$^{-1}$ \citep{Wyttenbach2020}. These measurements are compatible with a ``Balmer-driven'' evaporation, since WASP-33b orbits an A5 star, while KELT-9b orbits an A0V star, where the Balmer flux is extremely high.

\section{Conclusions}
We observed four transits of the ultra-hot Jupiter WASP-33b with the CARMENES and HARPS-N spectrographs. After the correction of the RM and CLV effects, we detected the Balmer H$\alpha$, H$\beta$, and H$\gamma$ transmission spectra of the planetary atmosphere.
The combined H$\alpha$ transmission spectrum has a large absorption depth of 0.99$\pm$0.05\,\%, indicating that the line probes neutral hydrogen atoms in the high-altitude thermosphere. Although the detection of the Balmer lines is unambiguous, the strengths of the lines are affected by the stellar pulsation. Future modelling and correction of the spectral pulsation feature will enable a better constrain the line strength.

We fitted the observed Balmer lines using the \texttt{PAWN} model assuming that the atmosphere is hydrodynamic and in LTE. The model fit returns a thermospheric temperature of $T=12200^{+1300}_{-1000}$ K and a mass-loss rate ${\rm \dot{M}}=10^{11.8^{+0.6}_{-0.5}}$ g\,s$^{-1}$. The high mass-loss rate is consistent with theoretical predictions for UHJs orbiting early type stars \citep[e.g.,][]{Fossati2018, Garcia-Munoz2019}.

The Balmer lines have so far been detected in five ultra-hot Jupiters (KELT-9b, KELT-20b/MASCARA-2b, WASP-12b, WASP-121b, and WASP-33b). Balmer absorption is probably a common spectral feature in the transmission spectra of ultra-hot Jupiters because their hot atmospheres are intensively irradiated by their host stars, which could produce a large number of hydrogen atoms in the excited state. However, for some UHJs, their low atmospheric scale heights \citep[see e.g. the case of WASP-189b,][]{Cauley2020} or the Rossiter-McLaughlin effect \citep[e.g.,][]{Casasayas-Barris2020} could hamper the detection of the Balmer features.
Extending the observations to a larger UHJ sample will enable a systematic study of the Balmer lines and the thermospheric conditions.

\begin{acknowledgements}
We thank the referee for the useful comments.
F.Y. acknowledges the support of the DFG priority program SPP 1992 "Exploring the Diversity of Extrasolar Planets (RE 1664/16-1)".
CARMENES is an instrument for the Centro Astron\'omico Hispano-Alem\'an (CAHA) at Calar Alto (Almer\'{\i}a, Spain), operated jointly by the Junta de Andaluc\'ia and the Instituto de Astrof\'isica de Andaluc\'ia (CSIC).
  
  CARMENES was funded by the Max-Planck-Gesellschaft (MPG), 
  the Consejo Superior de Investigaciones Cient\'{\i}ficas (CSIC),
  the Ministerio de Econom\'ia y Competitividad (MINECO) and the European Regional Development Fund (ERDF) through projects FICTS-2011-02, ICTS-2017-07-CAHA-4, and CAHA16-CE-3978, 
  and the members of the CARMENES Consortium 
  (Max-Planck-Institut f\"ur Astronomie,
  Instituto de Astrof\'{\i}sica de Andaluc\'{\i}a,
  Landessternwarte K\"onigstuhl,
  Institut de Ci\`encies de l'Espai,
  Institut f\"ur Astrophysik G\"ottingen,
  Universidad Complutense de Madrid,
  Th\"uringer Landessternwarte Tautenburg,
  Instituto de Astrof\'{\i}sica de Canarias,
  Hamburger Sternwarte,
  Centro de Astrobiolog\'{\i}a and
  Centro Astron\'omico Hispano-Alem\'an), 
  with additional contributions by the MINECO, 
  the Deutsche Forschungsgemeinschaft through the Major Research Instrumentation Programme and Research Unit FOR2544 ``Blue Planets around Red Stars'', 
  the Klaus Tschira Stiftung, 
  the states of Baden-W\"urttemberg and Niedersachsen, 
  and by the Junta de Andaluc\'{\i}a.
  
  Based on data from the CARMENES data archive at CAB (CSIC-INTA).
  
  We acknowledge financial support from the Agencia Estatal de Investigaci\'on of the Ministerio de Ciencia, Innovaci\'on y Universidades and the ERDF  through projects PID2019-109522GB-C51/2/3/4, PGC2018-098153-B-C33, AYA2016-79425-C3-1/2/3-P, ESP2016-80435-C2-1-R and the Centre of Excellence ``Severo Ochoa'' and ``Mar\'ia de Maeztu'' awards to the Instituto de Astrof\'isica de Canarias (SEV-2015-0548), Instituto de Astrof\'isica de Andaluc\'ia (SEV-2017-0709), and Centro de Astrobiolog\'ia (MDM-2017-0737), and the Generalitat de Catalunya/CERCA programme.  
  
A.W. acknowledges the financial support of the SNSF by grant number P400P2\_186765.
P.M. acknowledges support from the European Research Council under the Horizon 2020 Framework Program via ERC grant 832428.
I.S. acknowledges funding from the European Research Council (ERC) under the European Union's Horizon 2020 research and innovation program under grant agreement No 694513.
M.L. achkowledges the funding from the project ESP2017-87143-R.  
  
  This work is based on observations made with the Italian Telescopio Nazionale Galileo (TNG) operated on the island of La Palma by the Fundaci\'on Galileo Galilei of the INAF (Istituto Nazionale di Astrofisica) at the Spanish Observatorio del Roque de los Muchachos of the Instituto de Astrofisica de Canarias. 

\end{acknowledgements}

\bibliographystyle{aa} 

\bibliography{W33-Ha-refer}

\appendix
\section{Additional figures}

\begin{figure*}
\centering
\includegraphics[width=0.95\textwidth]{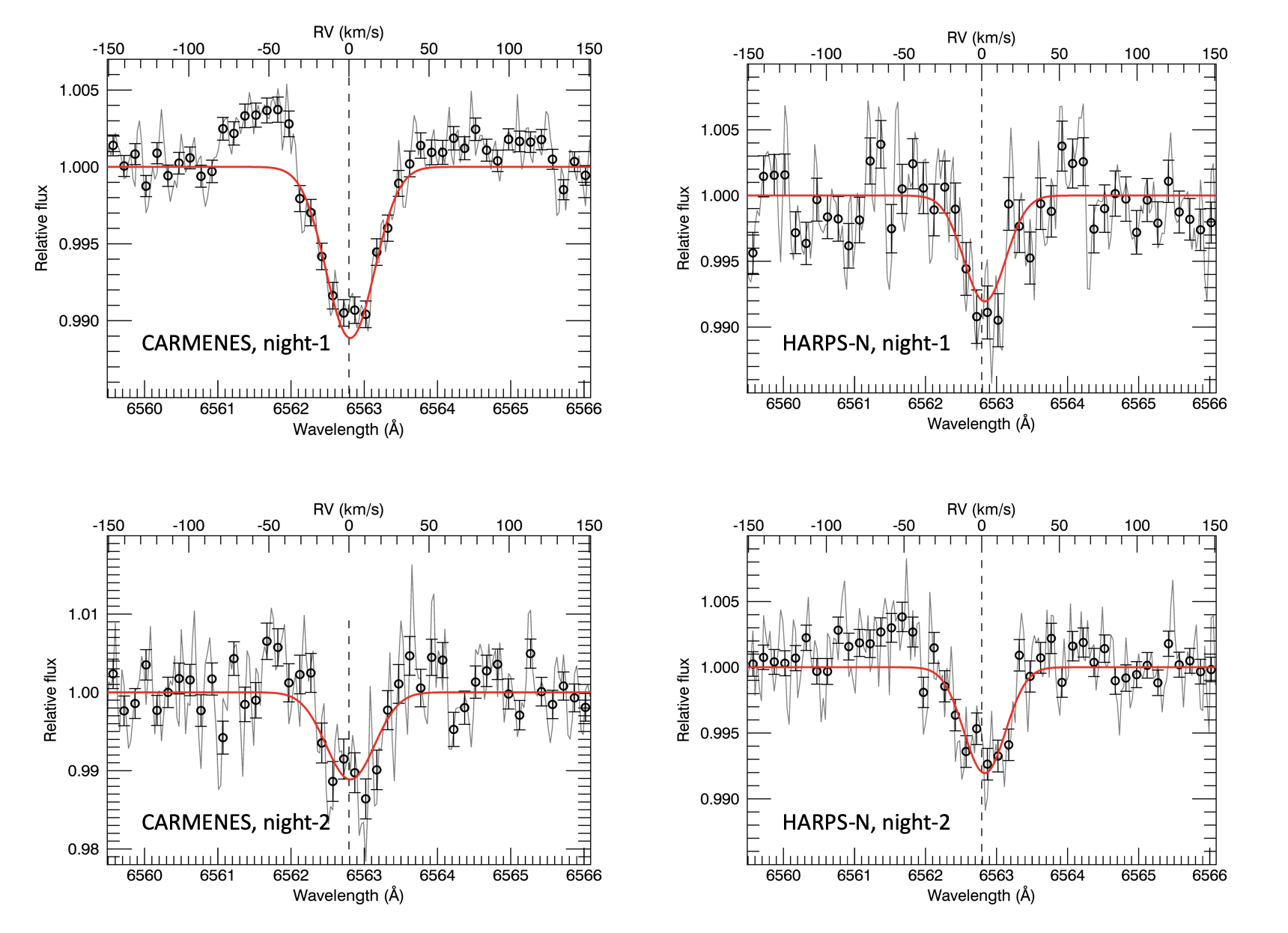}
  \caption{$\mathrm{H\alpha}$ transmission spectra of the four nights. The grey lines are the original spectra and the black points are spectra binned every five points. The red lines are the same best-fit Gaussian functions as in Fig. \ref{Spec-Ha}.
  }
     \label{App-Ha-individual}
\end{figure*}
\clearpage

\end{document}